\documentstyle[emulateapj,apjfonts,epsfig]{article}

\def\spose#1{\hbox to 0pt{#1\hss}}
\def\lax{$\mathrel{\spose{\lower 3pt\hbox{$\mathchar"218$}}
     \raise 2.0pt\hbox{$\mathchar"13C$}}$}
\def\gax{$\mathrel{\spose{\lower 3pt\hbox{$\mathchar"218$}}
     \raise 2.0pt\hbox{$\mathchar"13E$}}$}

\newcommand{\cyg}{V404\,Cyg}
\newcommand{\nmon}{A0620--00}
\newcommand{\gro}{GRO\,J1655--40}
\newcommand{\xte}{XTE\,J1550--564}
\newcommand{\chandra}{{\it Chandra}}
\newcommand{\asca}{{\it ASCA}}
\newcommand{\rosat}{{\it ROSAT}}
\newcommand{\sax}{{\it BeppoSAX}}
\newcommand{\xmm}{{\it XMM-Newton}}
\newcommand{\lum}{\thinspace\hbox{$\hbox{erg}\thinspace\hbox{s}^{-1}$}}
\newcommand{\flux}{\thinspace\hbox{$\hbox{erg}\thinspace\hbox{cm}^{-2}
            \thinspace\hbox{s}^{-1}$}}
\begin{document}

\title{The X-ray Spectra of Black Hole X-ray Novae in Quiescence as 
Measured
by {\it Chandra}}
\author{Albert K.H.~Kong, Jeffrey E.~McClintock, Michael R.~Garcia,
Stephen S.~Murray}
\affil{Harvard-Smithsonian Center for Astrophysics, 60
Garden Street, Cambridge, MA 02138}
\author{Didier Barret}
\affil{Centre d'Etude Spatiale des Rayonnements, 9 Avenue du Colonel
Roche, 31028 Toulouse, France}

\slugcomment{Accepted for publication in ApJ}

\begin{abstract}

We present {\it Chandra} observations of black hole X-ray novae \cyg,
\nmon, \gro\ and \xte\ in quiescence. Their quiescent spectra can be
well fitted by a power-law model with number slope $\alpha \sim 2$.
While a coronal (Raymond-Smith) model is also a statistically
acceptable representation of the spectra, the best fit temperatures of
these models is $\sim 5$ times higher than that seen in active stellar
coronae.  These four spectra of quiescent X-ray novae are all
consistent with that expected for
accretion via an advection-dominated accretion flow (ADAF) and
inconsistent with that expected from a stellar corona.  This evidence
for continued accretion in quiescence further strengthens the case for
the existence of event horizons in black holes.  Both \nmon\ and \gro\
were fainter than in previous observations, while \cyg\ was more
luminous and varied by a factor of 2 in a few ksec.  
A reanalysis of the X-ray data for \xte\ shows that (like \cyg\ and
\nmon) its luminosity exceeds the maximum prediction of the coronal
model by a large factor.  The 0.3--7 keV luminosity of the four
sources studied ranges from $\sim 10^{30}-10^{33}$ \lum.

\end{abstract}

\keywords{binaries: close --- black hole physics --- stars: individual
(\cyg, \nmon, \gro, \xte)
--- X-rays: stars}

\section{Introduction}

X-ray Novae (XN) are compact binary systems in which a
Roche-lobe-overflowing main sequence or subgiant star, typically $\sim
1\,M_\odot$, transfers matter onto a black hole (BH) or neutron star
(NS) primary (for a review, see van Paradijs \& McClintock 1995;
Tanaka \& Lewin 1995; Tanaka \& Shibazaki 1996). XN are highly
variable and undergo rare but dramatic X-ray and optical
outbursts. For most of the time, XN are in a quiescent state and are
very faint. During quiescence, the mass accretion rate from the disk
to the compact object may be very small, producing a low level
(perhaps no) X-ray emission. X-ray observations of quiescent XN have
been hindered due to the limited sensitivity of previous X-ray
telescopes. Nonetheless, several of the brightest black hole X-ray
novae (BHXN) have been detected with \rosat, \asca\ and \sax. This
quiescent X-ray (and associated non-stellar optical) emission is
difficult to explain using standard accretion disk models. Narayan,
McClintock, \& Yi (1996), Narayan, Barret, \& McClintock (1997a)
and Narayan, Garcia, \& McClintock (2001) showed that the observations
can be explained by an advection-dominated accretion flow (ADAF) model.

An ADAF is an accretion flow in which most of the energy is stored in
the accreting gas rather than being radiated away promptly, as in a
thin accretion disk.  This thermal energy is advected with the flow to
the center -- hence the name ADAF. If the accretor is a BH, the gas
with all its thermal energy will be lost from view as it falls through
the event horizon. However, in the case of a NS, the accretion energy
will eventually be radiated from the star's surface. This difference
can explain the fact that quiescent BHs are much fainter than
quiescent NSs.  Using pre-\chandra\ data, Narayan, Garcia, \&
McClintock (1997b) showed that BHs display a large variation of
luminosity between their bright and their faint states, while NSs have
a much smaller variation.  Menou et al. (1999) subsequently pointed
out that in comparing the luminosities of BH and NS systems, it is
important to compare systems with comparable orbital periods.  More
recently, Garcia et al. (2001; hereafter G01) presented a
comprehensive study of a series of \chandra\ observations of BHXN in
quiescence; they confirmed that the quiescent X-ray luminosities of
BHXN are $\sim 100$ lower than those of neutron star X-ray novae
(NSXN). Such findings provide strong evidence that BHs have event
horizons.

Recently, Bildsten \& Rutledge (2000) suggested that the rapidly
rotating secondaries of BHXN may generate stellar coronae with
sufficient X-ray luminosity to account for the observed quiescent
luminosities of many of these systems.  Based on an analogy to the
`saturated' coronae in the most luminous RS~CVn stars, the coronae in
quiescent BHXN are predicted to have maximum luminosities of 0.1\% of
the stellar bolometric luminosity, and X-ray spectra that are typical
of moderately hot ($kT$ \lax 1~keV), optically thin thermal plasmas.
While the X-ray luminosity of \cyg\, is too high to be produced by a
stellar corona, previous observations with modest sensitivity have
indicated that the luminosities of other BHXN are consistent with a
saturated corona (Bildsten \& Rutledge 2000).  For these systems, the
high S/N X-ray spectra attainable with {\it Chandra} and {\it
XMM-Newton} can provide a critical test of the possible coronal origin
of the quiescent X-ray luminosity (Bildsten \& Rutledge 2000; Lasota
2001).

In this paper, we report the detailed analysis of {\it Chandra}
spectra of the brightest three quiescent BHXN observed under an AO-1
GTO program (\cyg, \nmon\ and \gro). We also reanalyzed the spectrum
of a fourth BHXN (\xte) observed under a DDT proposal.  We note that
three other BHXN (GRO\,J0422+32, GS\,2000+25 and 4U\,1543--47)
observed under our AO-1 GTO and GO programs provided insufficient
counts for spectral analysis.  We briefly describe previous quiescent
observations of these four sources in \S\,2. In \S\,3 we outline our
analysis procedure and report the results in \S\,4. The results are
discussed in \S\,5.

\section{Previous Quiescent X-ray Observations}

All four BHXN have been observed previously in the X-ray, and a
summary of previous quiescent observations is given in Table 1; we
here discuss them briefly.

{\sl \cyg}~--- This relatively bright quiescent BHXN has previously
been observed by \rosat, \asca\ and \sax\ (see Table 1). In general,
the X-ray spectrum can be fitted by a power-law model with photon
index $\alpha\sim 2$ and $N_H\sim (1-2)\times 10^{22}$ cm$^{-2}$; the
luminosity is $\sim 10^{33}$ ergs s$^{-1}$ (Narayan et al. 1997a). We 
also
note that the quiescent source flux can vary
on short time scales. Wagner et al. (1994) reported that \cyg\
decreased in intensity by a factor of 10 in $<$ 0.5 day and showed
variability by a factor of $\sim 2$ on time scales of $\sim 30$
minutes.

{\sl A0620--00}~--- This source was observed by \rosat\ in 1992 during
its quiescent state (McClintock et al. 1995; Narayan et
al. 1997a). The $39\pm8$ counts detected allowed only a modest 
estimate of the source spectrum. Simple one component models fit the
spectrum equally well: for example a power-law with $\alpha\sim 3.5$
and $N_H=(0.1-1)\times 10^{22}$ cm$^{-2}$ or a blackbody with
$kT=0.16^{+0.10}_{-0.05}$ keV.  The luminosity is $\sim
5\times10^{30}$ ergs s$^{-1}$. An \asca\ observation in 1994 March
failed to detect the source; a $3\sigma$ upper limit on the luminosity
was $8\times 10^{30}$ ergs s$^{-1}$ (Asai et al. 1998).

{\sl GRO\,J1655--40}~--- The only quiescent observation of
GRO\,J1655--40 was taken in 1996 March with \asca\ (Ueda et al. 1998;
Asai et al. 1998). The spectrum can be fitted by a power-law model
with a photon index $\alpha\sim 0.7$ and $N_H < 3\times10^{21}$
cm$^{-2}$; the source luminosity is $3\times10^{32}$ ergs s$^{-1}$ in
0.5--10 keV. However, we note that this observation was taken between
two outbursts separated by $\sim 1$ years and therefore it may not
represent the true quiescent emission.

{\sl XTE\,J1550--564}~--- This microquasar system was observed as a
DDT program on 2000 August 21 and 2000 September 11, which were
$>$~120~d after the peak of the 2000 outburst of the source; a
detailed spectral analysis has already been given by Tomsick, Corbel,
\& Kaaret (2001). The energy spectrum can be fitted by an absorbed
power-law spectrum with $\alpha=2.3^{+0.41}_{-0.48}$ and
$N_H=(8.5^{+2.2}_{-2.4})\times10^{21}$ cm$^{-2}$; the mean luminosity
(0.5--7 keV) is about $6.7\times10^{32}$ \lum.

\begin{table*}
{\scriptsize
\begin{center}
TABLE~1

{\sc Previous Quiescent Observations of Black Hole X-ray Novae}

\vspace{1mm}
\begin{tabular}{llcccccc}
\hline \hline
Source&Date& Instrument & $N_H$& $\alpha$ & Luminosity & Distance 
&References\\
    &&            & ($10^{22}$ cm$^{-2}$) & &($10^{33}$ \lum)  & 
(kpc)&\\
\hline
\cyg\ &1992 Nov & \rosat & 2.29 $^a$ & 6 $^\dagger$ & 8.1 (0.1--2.4
keV)& 3.5 &1\\
&&  & 2.1 $^a$ & $4.0^{+1.9}_{-1.5}$ & 1.1 (0.7--2.4 keV)&&2 \\
&1994 May & \asca & $1.1^{+0.3}_{-0.4}$& $2.1^{+0.5}_{-0.3}$ &
1.20 (1--10 keV)& &3\\
&1996 Sept & \sax & 1.0 (fixed)& $1.9^{+0.6}_{-0.3}$  & 1.04 (1--10
keV) & &4\\
A0620--00 & 1992 Mar & \rosat\ &0.16 (fixed) & $3.5^{+0.8}_{-0.7}$
&0.004 (0.4--1.4 keV)& 1.0 & 2\\
&  1994 Mar & \asca\ &1.6 (fixed) & 2 (fixed) &$< 0.008$
(0.5--10 keV)& &5\\ 
GRO\,J1655--40&1996 Mar & \asca\ & $< 0.3$&$0.7^{+2.1}_{-0.4}$&0.3 
(0.5--10
keV)&3.2&5\\
\xte & 2000 Aug \& Sep & \chandra\ & $0.85^{+2.2}_{-2.4}$ &
$2.3\pm0.4$ & 0.67 (0.5--7 keV) & 2.5--6.3 & 6,7\\ 
\hline
\end{tabular}
\end{center}
\hspace{15mm}
\begin{minipage}[h]{7in}
NOTES --- $^\dagger$ Uncertainty not given. For \xte, the luminosity
is based on a distance of 4 kpc.\hfill\\
(1) Wagner et al. 1994; (2) Narayan et al. 1996; (3) Narayan et
al. 1997a; (4) Campana et al. 2001; (5) Asai et al. 1998; (6) Tomsick
et al. 2001; (7) Orosz et al. 2001
\end{minipage}
}
\end{table*}

\section{Chandra Observations and Data Reduction}

{\bf \cyg}~--- \chandra\ observed \cyg\ on 2000 April 26 for a total
of 10,295 s. Our observations cover spectroscopic phases 0.44--0.46
(Casares \& Charles 1994), where phase zero corresponds to the closest
approach of the secondary star. The source was positioned on the
ACIS-S3 CCD with an offset of $40''$ from the nominal pointing for
the S3. The data were collected using a 1/4 subarray mode, which
boosted the time resolution to 1.14~s.  The CCD temperature was
$-120^{\circ}$C.  Standard pipeline processed level 2 data were used for 
the
analysis. \cyg\ was clearly detected and the source position is
$\alpha=$ 20h\,24m\,03.82s, $\delta=$ +33d\,52m\,02.14s (J2000), which
is in good agreement with the optical and radio position of \cyg\
(Wagner et al.  1991).

The \chandra\ detectors are known to experience periods of high
background, which are particularly significant for the S3 chip
(e.g. Garcia et al. 2000).  We searched for such background flares in
our data by examining the light curve of the entire S3 chip minus the
source regions.  We found that the background was very stable during
the whole observation with an average count rate of 0.13 count
s$^{-1}$. In order to reduce the background, we only analyzed data
from 0.3--7 keV.  We extracted data from a circle of 3 pixels ($\sim
1.5''$) centered on \cyg\ and background from an annulus with inner
and outer radii of 10 and 50 pixels, respectively. There were 1587
counts in the source region and the expected number of background
counts in the source region was only 0.4~counts.

{\bf \nmon}~--- This source was observed by \chandra\ on 2000 February
29 for 44,000 s. ACIS-S was operated in the standard configuration
with a time resolution of 3.24 s. \nmon\ was observed on the S3 chip
with a $40''$ offset from the nominal pointing. Background was
examined; only intervals where the source-free count rate was less
than 0.15 count s$^{-1}$ were selected for analysis. The total net
exposure time is 41,189 s. The source position is $\alpha=$
06h\,22m\,44.48s, $\delta=$ -00d\,20m\,46.36s (J2000) which is
consistent with the optical position (Liu, van
Paradijs, \& van den Heuvel 2001). The observations cover
spectroscopic phases 0.09--1.67 (Orosz et al. 1994; Leibowitz, Hemar,
\& Orio 1998). Only data from 0.3--7 keV were used for spectral
analysis. We extracted data from a circle of $1.86''$ centered on
\nmon. This relatively large aperture encompasses all of the counts in
the central region that might reasonably be attributed to the
source. There were 137 counts in the source region.  The background
counts in a $1.86''$ aperture are estimated to be 1.2. This small
background level was not subtracted.

{\bf \gro}~--- \chandra\ observed \gro\ on 2000 July 1 for 43,000 s,
which corresponds to spectroscopic phases of 0.49--0.68 (van der Hooft
et al. 1998). The source was located on ACIS-S3 with $40''$ offset
from the aim-point; standard 3.24 s frame transfer time was
employed. Good data were selected with background count rate $< 0.15$
count s$^{-1}$, resulting in a net exposure of 42,506 s. \gro\ was
very faint; by filtering the data from 0.3--7 keV and applying a
circular extraction region of $1.41''$ centered on the source, only 66
counts were collected. This choice of aperture encompasses all of
counts in the central region that are attributable to the source. The
estimated background counts in a $1.41''$ aperture is estimated to be
0.7; this background was not subtracted.  The Chandra source position
is $\alpha=$ 16h\,54m\,00.09s, $\delta=$ -39d\,50m\,45.37s (J2000),
which is consistent with the radio and optical position (Hjellming
1994; Bailyn et al. 1995).

{\bf \xte}~--- The source was observed on 2000 August 21 for
$\sim$~5,000~s and 2000 September 11 for an additional $\sim$ 5,000 s;
the observations cover spectroscopic phases 0.06--0.11 and 0.63--0.68,
respectively (Orosz et al. 2001). Technical details of the
observations can be found in Tomsick et al. (2001). We
used similar procedures to those outlined in Tomsick et al. (2001) to 
reduce
the data. However, we extracted data from 0.3--7 keV and used a
smaller circular extraction region with a radius of $2''$, which is
sufficient to encompass all of the counts in the central region. There
are 66 and 109 counts in the first and the second observations,
respectively; we ignored the background counts in the source region,
which we estimated to be 0.2 counts for the first observation and 0.3
counts for the second observation .

\section{Spectral Analysis}

\subsection{\cyg}

Spectra were extracted with CIAO v2.1
\footnote{http://asc.harvard.edu/ciao/} and were analyzed with XSPEC v11
\footnote{http://heasarc.gsfc.nasa.gov/docs/xanadu/xspec/index.html}
and also SHERPA v2.1.2
\footnote{http://asc.harvard.edu/ciao/download/doc/sherpa\_html\_manual/index.html}.
The results from both analysis systems were consistent, and we report
the XSPEC results herein.  In order to allow $\chi^2$ statistics to be
used, all the spectra were grouped into at least 30 counts per
spectral bin. Response files were selected according to the CCD
temperature with standard CIAO routines. We fit the data with several
single-component spectral models including power-law, thermal
bremsstrahlung, Raymond-Smith and blackbody models including
interstellar absorption.  The best-fit parameters determined by these
fits are shown in Table 2.

All models except the blackbody model gave statistically acceptable
fits to the data ($\chi^2/\nu$ \lax 1).  The power-law model provides
the best fit, and yields parameters consistent with previous
observations (e.g. $\alpha = 1.81 \pm 0.14$; see Table 1). This best
fitting model is shown in Figure 1 and the corresponding plot of
confidence regions for column density ($N_H$) and photon index
($\alpha$) are shown in Figure 2a.  The confidence bounds for the
Raymond-Smith model are shown in Figure 2b.  The best fit temperature
for this model is $kT = 7.5$~keV, and the 90\% lower limit on the
temperature is $kT > 6.1$~keV.

\begin{table*}[t]
{\footnotesize
\begin{center}
TABLE~2

{\sc Best-fitting Spectral Parameters}

\vspace{1mm}
\begin{tabular}{llcccccc}
\hline \hline
Source& Model & $N_H$& $\alpha$ & $kT/kT_{RS}$\,$^a$ & 
$\chi^2_{\nu}/dof$ (prob)& CASH &Flux$^c$\\
      && ($10^{21}$ cm$^{-2}$)& & (keV) &  & M-C Prob$^b$ &\\
\hline
\cyg\ &Power-law & $6.98\pm0.76$& $1.81\pm0.14$& & 0.92/45 (0.63) & &
1.42\\
& Bremsstrahlung & $6.04^{+0.60}_{-0.55}$& & $6.68^{+2.49}_{-1.50}$& 
0.94/45 (0.57)&
      &1.40\\
&Raymond-Smith&$5.82^{+0.56}_{-0.50}$& &$7.54^{+2.70}_{-1.43}$ &1.11/45 
(0.28)&
      & 1.57\\
&Blackbody & $2.30\pm0.42$& & $0.81\pm0.04$ & 2.09/45 (0.00002) && 1.26\\
&&&&&&\\
&Power-law & 5.40 (fixed) &$1.55\pm0.07$& & 1.20/46 (0.17) & &1.47\\
& Bremsstrahlung & 5.40 (fixed)& & $8.66\pm2.13$& 1.0/46 (0.46)&
      &1.42\\
&Raymond-Smith& 5.40 (fixed)& &$8.89\pm1.57$  &1.13/46 (0.25)&
      & 1.57\\
&Blackbody & 5.40 (fixed) & & $0.69\pm0.03$  & 3.49/46 ($10^{-14}$) && 
1.15\\
&&&&&&\\
\nmon & Power-law & $2.37^{+1.14}_{-1.04}$ & $2.19\pm0.50$ & & 0.71/11
(0.73)& 0.78&0.018\\
       &  Bremsstrahlung& $1.52^{+0.72}_{-0.67}$ &
&$3.11^{+3.59}_{-1.17}$&0.75/11 (0.69)& 0.74&0.018\\
       & Raymond-Smith& $1.05^{+0.57}_{-0.50}$ & &
$5.46^{+6.51}_{-2.07}$&1.03/11 (0.42)&0.48&0.022\\
       & Blackbody & 0\,$^d$ & & $0.57^{+0.06}_{-0.07}$ &1.58/11 (0.10)& 
0.10&0.017\\
&&&&&&\\
       & Power-law & $1.94\pm0.28$ (fixed)& $2.07^{+0.28}_{-0.19}$ & &
0.71/12 (0.74)&0.75&0.018\\
       &  Bremsstrahlung& $1.94\pm0.28$ (fixed)&
&$2.55^{+1.44}_{-0.73}$&0.78/12 (0.67)&0.68&0.016\\
       & Raymond-Smith& $1.94\pm0.28$ (fixed)&
&$4.15^{+2.66}_{-1.30}$&1.38/12 (0.17)&0.14&0.023\\
       &Blackbody & $1.94\pm0.28$ (fixed) & & $0.30\pm0.03$ &2.39/12 
(0.004)& 0.00& 0.009\\
&&&&&&\\ 
&&&&&&\\      
\gro & Power-law & $8.59_{-4.52}^{+6.19}$ &$1.70_{-0.78}^{+0.88}$ &
&0.83/9 (0.59)&0.66&0.017\\
      &
Bremsstrahlung&$7.72^{+5.11}_{-3.46}$&&$8.40^{+\infty}_{-5.73}$&0.83/9
(0.58)&0.66& 0.016\\
      &  Raymond-Smith &$7.18^{+4.23}_{-2.97}$&
&$12.24^{+\infty}_{-8.61}$ & 0.85/9 (0.56)&0.63&0.019\\      
& Blackbody &$3.03^{+3.47}_{-2.13}$&&$0.88^{+0.29}_{-0.18}$&0.94/9 
(0.49)&0.57&0.012\\
&&&&&&\\
      & Power-law & $6.66\pm0.57$ (fixed) &$1.47\pm0.40$ & &0.75/10 
(0.67)&0.60&0.016\\
      &  Bremsstrahlung& $6.66\pm0.57$ (fixed)
&&$13.21^{+\infty}_{-8.98}$&0.75/10 (0.68)&0.64&0.015\\
      & Raymond-Smith & $6.66\pm0.57$ (fixed) &&$17.15^{+\infty}_{-11.35
}$&0.77
/10 (0.65)&0.62&0.018\\
      & Blackbody & $6.66\pm0.57$ (fixed) &&
$0.76^{+0.14}_{-0.12}$&1.07/10 (0.38)&0.39&0.012\\
&&&&&&\\
\xte & Power-law & $8.73^{+2.42}_{-2.93}$ &
$2.28^{+0.47}_{-0.64}$ &&1.27/13 (0.22)& 0.22&0.16 \\
&  Bremsstrahlung& $6.93^{+2.13}_{-1.85}$ && $3.36^{+4.75}_{-1.33}$
&1.26/13 (0.23)& 0.24&0.15\\
& Raymond-Smith & $6.50^{+1.97}_{-1.62}$ &&$4.38^{+4.31}_{-1.57}$
&1.22/13 (0.25)& 0.21&0.18\\
& Blackbody & $3.04^{+1.80}_{-1.49}$ && $0.69^{+0.11}_{-0.09}$
&1.39/13 (0.16)&0.18&0.14\\

&&&&&&\\
& Power-law & $3.90\pm0.60$ (fixed) & $1.35\pm0.25$ &&1.68/14 (0.05)& 
0.02&0.17\\
&  Bremsstrahlung & $3.90\pm0.60$ (fixed) && $12.56^{+\infty}_{-7.18}$
&1.59/14 (0.07)& 0.04&0.15\\
& Raymond-Smith & $3.90\pm0.60$ (fixed) && $10.31^{+\infty}_{-5.16}$
&1.61/14 (0.07)& 0.03&0.16\\
& Blackbody & $3.90\pm0.60$ (fixed) && $0.65^{+0.08}_{-0.06}$ &1.39/14
(0.15)& 0.15&0.12\\

\hline
\end{tabular}
\end{center}
\hspace{22mm}
\begin{minipage}[h]{6in}
NOTES --- All quoted uncertainties are 90\% confidence.\\
Except for \cyg, the best-fit parameters and uncertainties are based
on the CASH statistic. The reduced $\chi^2$ values were obtained in a 
separate analysis using the $\chi^2$ statistic. \\
$^a$ Thermal  bremsstrahlung, blackbody or Raymond-Smith temperature 
(solar
abundance)\hfill\\
$^b$ For \nmon, \gro, and \xte, we list one minus the probability that the best fit model would produce a lower value of the CASH statistic than that calculated from the data, as determined via XSPEC Monte-Carlo simulations. A low entry indicates a poor fit.\hfill\\
$^c$ Absorbed flux in 0.3--7 keV (10$^{-12}$\flux)\hfill\\
$^d$ $N_H$ hit the minimum value of 0 allowed by XSPEC\hfill\\
\end{minipage}
}
\end{table*}

The hydrogen column density for \cyg\ from optical observations was
estimated to be $5.4\times 10^{21}$ cm$^{-2}$ ($A_V=3.1$; Casares \&
Charles 1994).  The best fit values for $N_H$ from the power law and
bremsstrahlung models are marginally higher than the
optically determined value, but this does not necessarily argue
against these models.  X-ray binaries often show absorption in the
X-ray flux which is somewhat higher than that determined by their
optical absorption (Garcia 1994; Vrtilek et al. 1991).

In order to test if the optically-determined absorption yields an
acceptable X-ray spectral fit, we re-ran the fits with the absorption
fixed to this value. The results of these fits are also given in
Table~2.  Even though this $N_H$ value is outside the 99\% confidence
bounds shown in Figure 2, these fits do yield acceptable values of
$\chi^2/\nu$ (except for the blackbody model). This is a reflection of
the fact that the minimum value of $\chi^2/\nu$ obtained with $N_H$ as
a free parameter is slightly less than one, thereby allowing points
outside the $\chi^2_{min} + 9.21$ (Lampton, Margon, \& Bowyer 1976)
contour to have $\chi^2/\nu \sim 1$.  For these fits with $N_H$ fixed,
the best fit temperature for the Raymond-Smith model is raised to
8.9~keV, and the 90\% lower limit is raised to $>7.2$~keV.

\vspace{5mm}
\vbox{
\centerline{
\psfig{figure=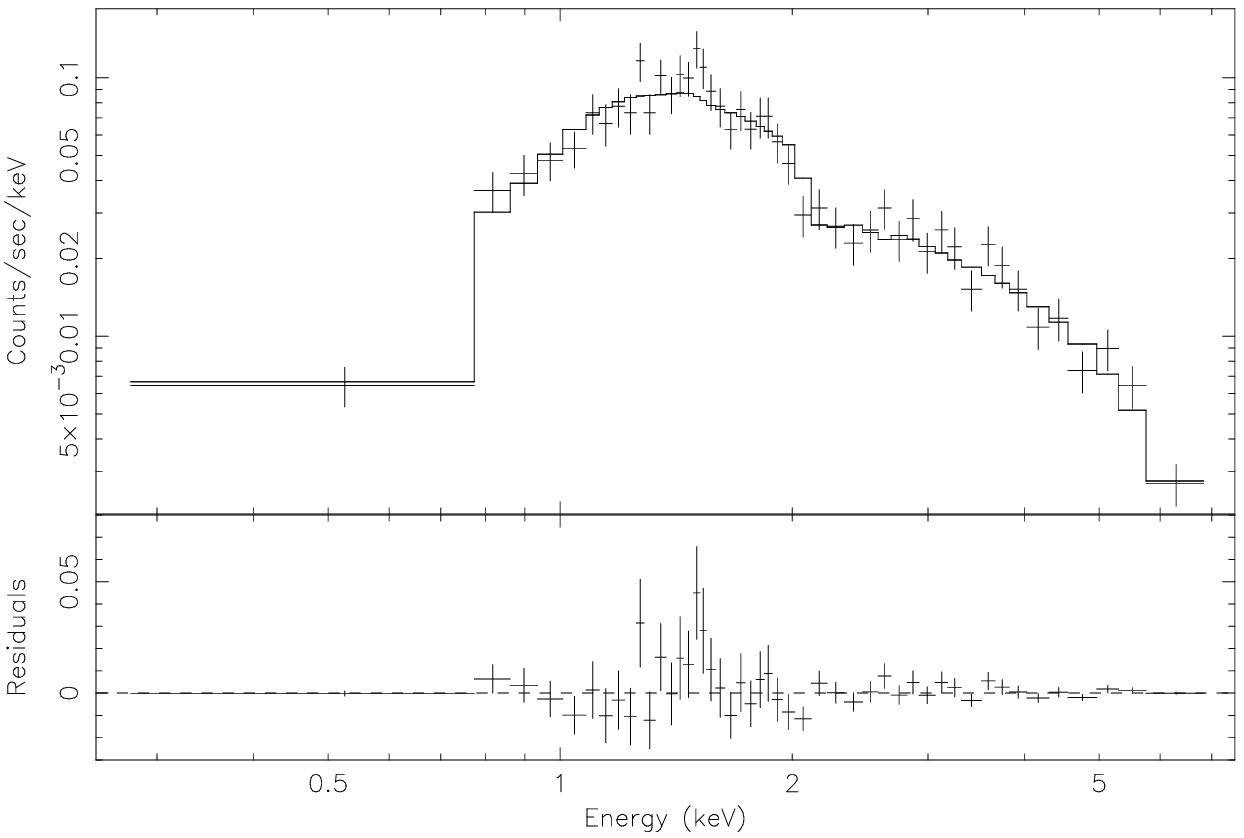,width=3.5in}
}
\centerline{
\begin{minipage}[h]{3.5in}
{\small Figure 1: {Upper panel: The \chandra\ spectrum of \cyg\ with an absorbed 
power-law
model ($\alpha=1.81$ and $N_H=6.98\times 10^{21}$ cm$^{-2}$). Lower
panel: residuals after subtracting the fit from the data in units of
$1\sigma$.\\ }}
\end{minipage}
}
}

We do not see any significant Fe-K line emission between 6.4--7 keV,
with a 90\% confidence upper limit of $\sim$ 800 eV (line width fixed
at 0.1 keV) on the equivalent
width.

\begin{figure*}[t]
\begin{center}
\psfig{file=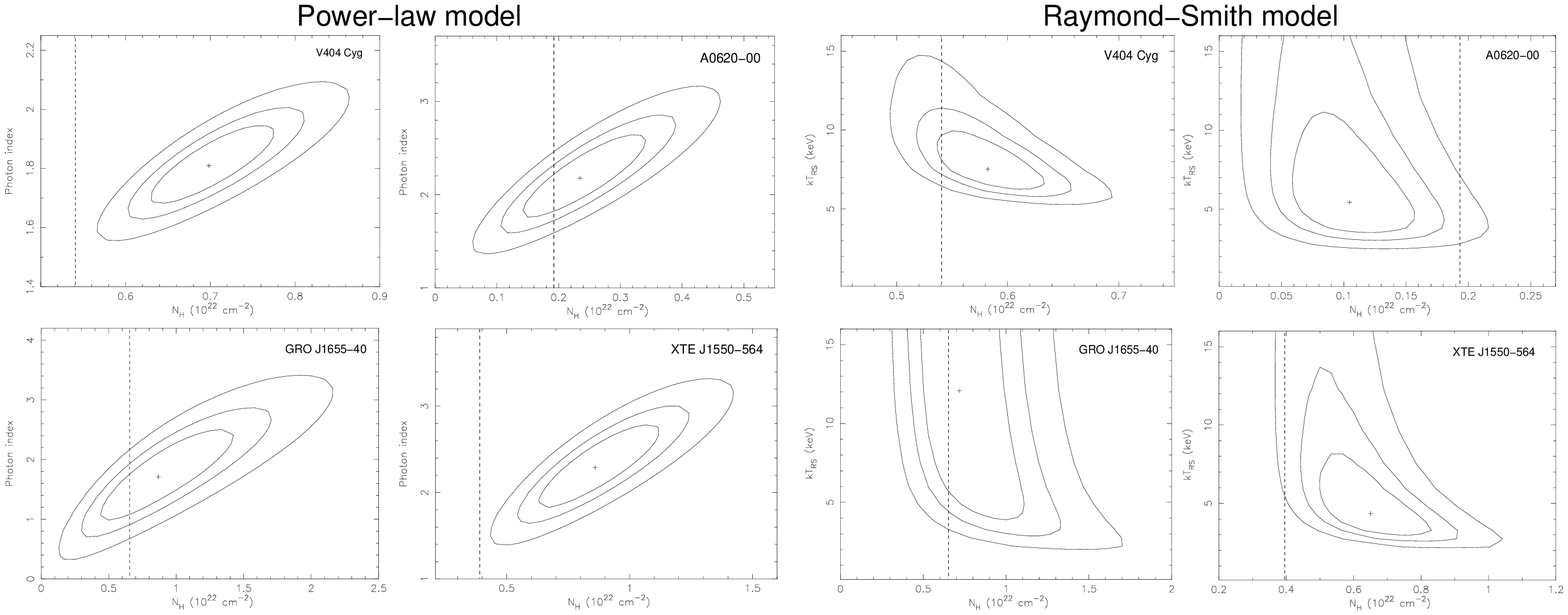,height=10cm,width=16.5cm}
\end{center}
{\small Figure 2: {Left: Contour plot for the column density ($N_H$) and photon
index ($\alpha$) derived from the {\it Chandra} spectrum of \cyg,
\nmon, \gro\ and \xte. The cross in the center marks the best fit
parameters and the contours encompass the 68\%, 90\% and 99\%
confidence levels. Vertical dashed lines show the optically determined
$N_H$.  Right: Contour plot for the column density ($N_H$) and
Raymond-Smith temperature ($kT_{RS}$) derived from the {\it Chandra}
spectrum of \cyg, \nmon, \gro\ and \xte. The cross in the center marks
the best fit parameters and the contours encompass the 68\%, 90\% and
99\% confidence levels. Except for \cyg, all of the plots were derived using CASH statistics.  Vertical dashed lines show the optically determined $N_H$.}}
\end{figure*}

\subsection{A0620--00}

We analyzed the energy spectrum of \nmon\ using procedures similar
to those discussed above for \cyg.  We grouped the data into
spectral bins containing at least 10 counts and used both $\chi^2$ and
CASH (Cash 1979) statistics to estimate the best-fit parameters and
their errors.  We chose to bin the data in order to achieve enough
counts per bin to employ the $\chi^2$ statistic.  However, binning the
data heavily can result in a loss of spectral information.  One can
also apply the Gehrels' approximation (Gehrels 1986) to permit the use
of fewer counts ($\leq 5$) per bin, but this approach over-estimates
the errors.  The CASH statistic is a maximum likelihood method
designed to estimate the best-fit parameters using unbinned or slightly binned data.
This is particularly useful when the source yields only very few
photons.  The disadvantage of the CASH statistic, relative to the
$\chi^2$ statistic, is that it does not provide a goodness-of-fit
criterion for comparing different models.  It is therefore worthwhile
to examine the results obtained using both $\chi^2$ and CASH
statistics.  In Table 2, except for \cyg, all the best-fit parameters
and errors are based on the CASH statistic on binned data; the reduced $\chi^2$
values are also shown to indicate the quality of the fit. In order to justify the significance of the CASH statistic, we performed Monte-Carlo simulations to estimate the significance level of the fits; these results are also given in Table 2. 

Both methods give very consistent results. We also ran the fits with unbinned data using the CASH statistic, and the results were consistent. We employed the same four
single-component models with interstellar absorption that we used for
\cyg; the best-fit parameters for the various spectral models are
shown in Table 2. Among all the models, the power-law gives the best
fit ($\chi^2/\nu = 0.71$, $\alpha=2.2 \pm 0.5$), while the blackbody
gives the worst fit ($\chi^2/\nu = 1.58$); Monte-Carlo simulations based on the CASH statistic also show similar results. The confidence regions for the power-law fit are shown in Figure 2a, and those for the
Raymond-Smith fit are shown in Figure 2b.  The best fit Raymond-Smith
temperature is $kT = 5.5$~keV, and the 90\% lower bound on the
temperature is $kT > 3.5$~keV.
 
The values of $N_H$ determined by the power-law and bremsstrahlung
fits are consistent with the optical value, corresponding to
$N_H=(1.94\pm0.28) \times 10^{21}$ cm$^{-2}$ (Wu et al. 1976, 1983;
Predehl \& Schmitt 1995).  The value of $N_H$ determined 
by the Raymond-Smith and blackbody models is lower than the optical
value.  This conclusion provides marginal evidence that neither the
blackbody nor the Raymond-Smith models are correct descriptions of the
source spectrum because X-ray fits tend to find $N_H$ higher than (or
consistent with) the optically determined value.  As in the case of
\cyg, we re-ran the fits with $N_H$ fixed at the optically determined
value.  The results of these fits are also shown in Table 2. The
derived parameters are consistent (within 1$\sigma$) with the results
obtained by varying $N_H$, except for the case of the blackbody model.
The best-fit temperature for the Raymond-Smith model is 4.1~keV, and
the 90\% lower bound is $>2.8$~keV.

Previously, the best measurement of the X-ray spectrum of \nmon\ was
that afforded by \rosat\ (Narayan et al. 1997a), which
gave $\alpha = 3.5 \pm 0.7$ with $N_H$ fixed to the optical value.
This led to the speculation that the quiescent X-ray spectra of BHXN
with orbital periods \lax~1~day might be softer than the spectra of
longer period systems.  However, this result was based on only $39\pm
8$~detected source photons in the presence of a significant
background.  The present result is much more robust because it is
based on more than 3 times as many counts, a negligible background,
and a much wider energy band.  It is important to note that A0620-00
was also a factor of $\sim 2$ fainter in this \chandra\ observation
than it was during the previous \rosat\ observation. The best fitting
power-law model indicates a 0.4--2.4~keV emitted flux of
$1.9\times10^{-14}$\flux, corresponding to a luminosity of 
$2.1\times10^{30}$
\lum, which is a factor of two below the \rosat\ value (see Table 1).

\subsection{GRO\,J1655--40}

The spectrum of \gro\ was analyzed using the same methods discussed
above for \nmon. The energy spectrum was grouped into spectral bins
containing at least 5 counts and fit using $\chi^2$ and CASH statistics.
Unbinned data was also fit using CASH statistic, and the results were
consistent.  All simple models give acceptable fits.  While the
blackbody model gives the poorest fits, it cannot be rejected on the
basis of $\chi^2/\nu$ and Monte-Carlo simulations.  However, the $N_H$ for the blackbody model is
slightly lower ($1.5 \sigma$) than the optical value of
$(6.66\pm0.57)\times 10^{21}$ cm$^{-2}$ (Predehl \& Schmitt 1995;
Hynes et al. 1998), while the other three models indicate values of
$N_H$ consistent with the optically-derived value. The relatively low
value of $N_H$ suggests that the blackbody model may not be a true
representation of the source spectrum.

The best fit temperature for the Raymond-Smith model is $kT=
12.24$~keV, and the 90\% lower limit on the temperature is $kT >
3.63$.  If we fix $N_H$ to the optical value, these values are raised
to $kT=17.15$~keV and $kT > 5.8$~keV.

As above, we list the best fit parameters in Table 2, and show a plot
of the confidence regions for power-law and Raymond-Smith fits in
Figures 2a and 2b.  It is important to note that these observations
show \gro\ to be a factor of $\sim 10$ fainter than previous quiescent
observations (see Table 1).  The observed 0.4--2.4~keV emitted flux,
for the best fitting power-law model, is $1.5 \times 10^{-14}$ \flux;
the observed 0.3--7.0~keV luminosity is $2.4 \times 10^{31}$ \lum .
The large decrease in flux and luminosity indicate that the previous
\asca\ observations may not have been taken during the true quiescent
state because the observations occurred between two outbursts.

\vspace{3mm}

\subsection{\xte}

We combined the two spectra of \xte\ as shown in Tomsick et
al. (2001), grouped the resulting data into bins containing at 
least 10 counts each, and fit the data to models using $\chi^2$ and CASH
statistics.  The results of the spectral fits are shown in Table 2,
and the corresponding parameter confidence regions are shown in Figure
2.  All four models yield statistically acceptable fits, and we see no
straightforward way to select one model over the others. With the
exception of the blackbody model, all of the models indicate that
$N_H$ is somewhat higher than that determined optically
(S$\acute{a}$nchez-Fern$\acute{a}$ndex et al. 1999).  However, as
indicated above, this is only a weak argument against the blackbody
model.  Fits with $N_H$ fixed to the optical value are also
statistically acceptable, and indicate harder ($\alpha$ lower, $kT$
higher) spectra than the fits with $N_H$ free.

The Raymond-Smith fits indicate a best fit temperature of $kT =
4.38$~keV, and a 90\% lower limit to the temperature of $kT >
2.81$~keV.  Fits with $N_H$ fixed to the optical value raise these
values to $kT = 10.31$~keV and $kT > 5.15$~keV.  The results of
power-law fit are consistent with those found by Tomsick et al. (2001).

In order to determine if the quiescent X-ray emission of \xte\ has a
flux consistent with a stellar corona, we calculated the unabsorbed
X-ray flux ($F_X$) and bolometric flux for \xte\ using the methods of
Bildsten \& Rutledge (2000). Based on our best-fit power-law result,
the unabsorbed 0.4--2.4 keV flux of \xte\ is $2.98\times 10^{-13}$
\flux . For the bolometric flux, we used
$F_{bol}=10^{-0.4(V_q+11.51+B.C.-A_V)}$ \flux (Bildsten \& Rutledge
2000), where B.C. is the bolometric correction for spectral type,
$V_q$ is the quiescent magnitude, and $A_V$ is the reddening. We
adopted a $V_q$ of $22\pm0.2$ and a spectral type of K3III from recent
VLT observations (Orosz et al. 2001), which indicates B.C.$=-0.8$. We
computed $F_{bol}$ using the $A_V$ determined from optical
observations ($A_V=2.17$; S$\acute{a}$nchez-Fern$\acute{a}$ndez et
al. 1999) and find $F_{bol} = 5.1 \times 10^{-13}$ \flux .  We also
determined $F_{bol}$ using the $A_V$ estimated from our X-ray spectral
fitting. For a power-law model, $N_H=8.73\times 10^{21}$ cm$^{-2}$
implies that $A_V=4.88$ (Predehl \& Schmitt (1995)], implying $F_{bol}
= 6.1 \times 10^{-12}$ \flux .  These values of $F_{bol}$ are
discussed in Section~6.

\section{Time-resolved Spectrum of \cyg}

The background-subtracted light curve of \cyg\  during
our observations is shown in Figure~3. The light curve shows a factor
of $\sim 2$ variability in a few ksec. We do not find any significant
peak in the power spectrum on timescales from 2.3\,s to 10,000\,s and
the $3\sigma$ upper limit on the semi-amplitude is 39\% (0.3--7 keV). 

The marked variability led us to search for spectral changes at
differing flux levels.  The data was divided into seven segments based
on the source intensity (see Figure 3).  The spectrum from each
segment contains at least 100 counts.  The results of fitting each
spectrum with a power-law model are shown in Table~3.  The best-fit
column density varied between $(2.91-11.08)\times 10^{21}$ cm$^{-2}$,
and the best fit photon index $\alpha$ varied between 1.1--2.4.  We
found no correlation between either the column density or $\alpha$ and
the flux.  However, we do find a positive correlation between the
absorption column and the photon index (see Figure 4) with a
correlation coefficient of 0.93 ($>$ 99\%).

\vbox{
\centerline{
\psfig{figure=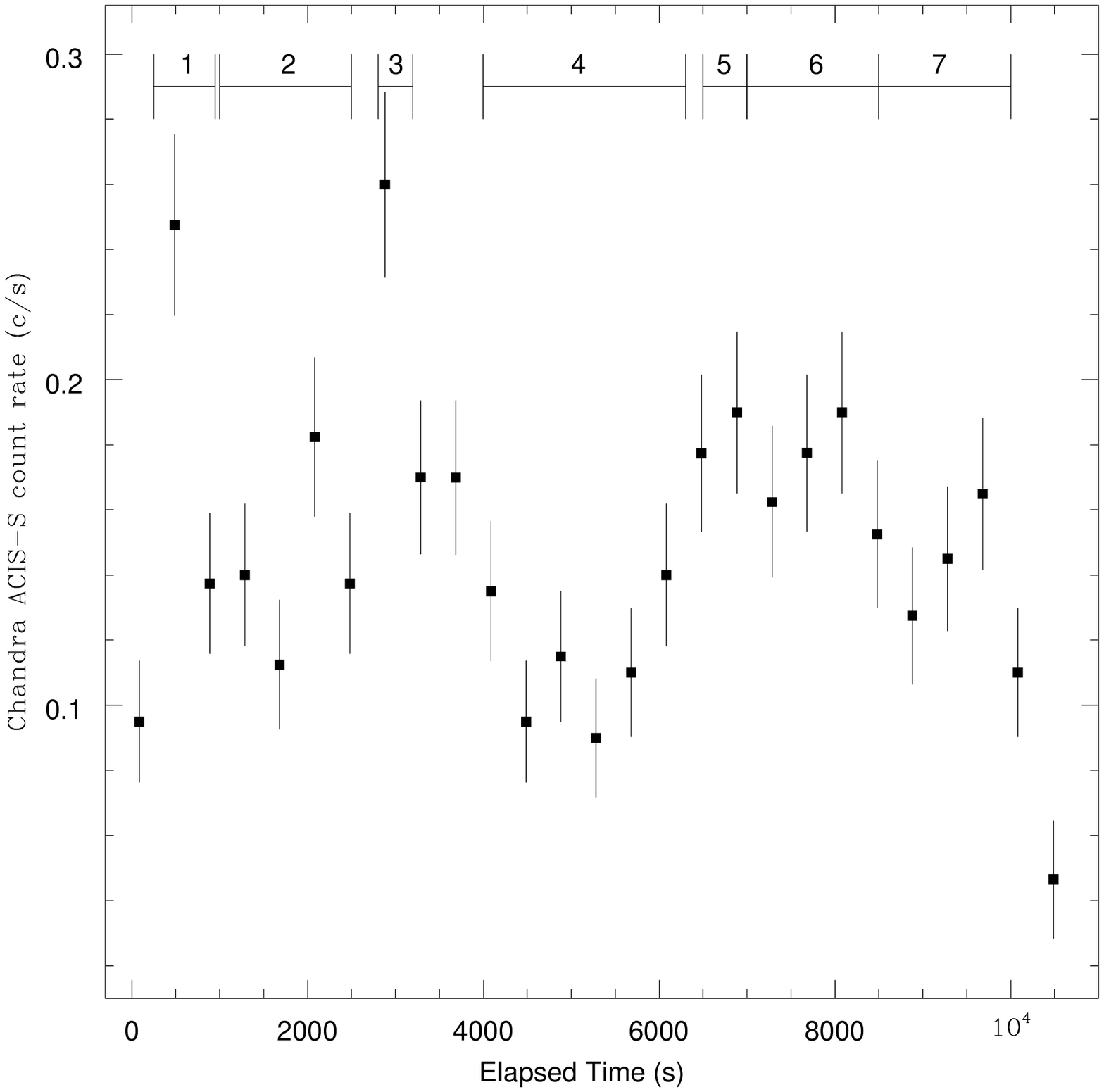,width=3.1in,height=2.7in}
}
\centerline{
\begin{minipage}[h]{3.5in}
{\small Figure 3: {\chandra\ ACIS-S 10\,ksec light curve of \cyg\ in the
0.3--7.0~keV band. The time
resolution is 500\,s. Also shown are the seven time intervals used for
time-resolved spectral analysis. \\}}
\end{minipage}
}}

However, we suspect that this correlation is not intrinsic to the
source, but is rather an artifact of the fitting process which links
$\alpha$ and $N_H$.  For example, we note that the slope of the
correlation is nearly (within $\sim 5$\%) the same as the slope of the
major axis of the parameter confidence contours (Figure 2a). Also, we
extracted and examined two spectra, one for count rates below
0.11~counts~s$^{-1}$ and the other for count rates above
0.18~counts~s$^{-1}$ (see Figure 3), and found them to be
identical. We conclude that the spectral shape does not vary with
intensity.

\vspace{-5mm}
\small
\begin{center}
TABLE~3

{\sc Time-resolved Spectral Parameters}

\begin{tabular}{lcccc}
\hline \hline
 & $N_H$& $\alpha$ & $\chi^2_{\nu}/dof$ & Luminosity$^a$\\
      & ($10^{21}$ cm$^{-2}$)& & \\
\hline
1 & $10.21\pm2.28$&$2.41\pm0.41$ & 1.21/11 & 8.07\\
2 & $5.08\pm1.19$&$1.57\pm0.26$&0.88/16& 2.81\\
3 & $6.14\pm1.97$&$1.72\pm0.40$&0.41/7 & 4.83\\
4 & $6.47\pm1.09$&$1.60\pm0.20$&0.85/28& 2.81\\
5 & $2.91\pm1.29$&$1.14\pm0.40$&1.17/6 & 4.26\\
6 & $5.86\pm1.18$&$1.56\pm0.24$&0.85/21& 3.82\\
7 & $11.08\pm2.15$&$2.22\pm0.31$&1.11/17& 5.32\\
\hline
\multicolumn{5}{l}{NOTES --- $^a$ Luminosity in 0.3--7 keV 
(10$^{33}$\lum),}\\
\multicolumn{5}{l}{assuming a distance of 3.5 kpc}
\end{tabular}
\end{center}

\vbox{
\centerline{
\psfig{figure=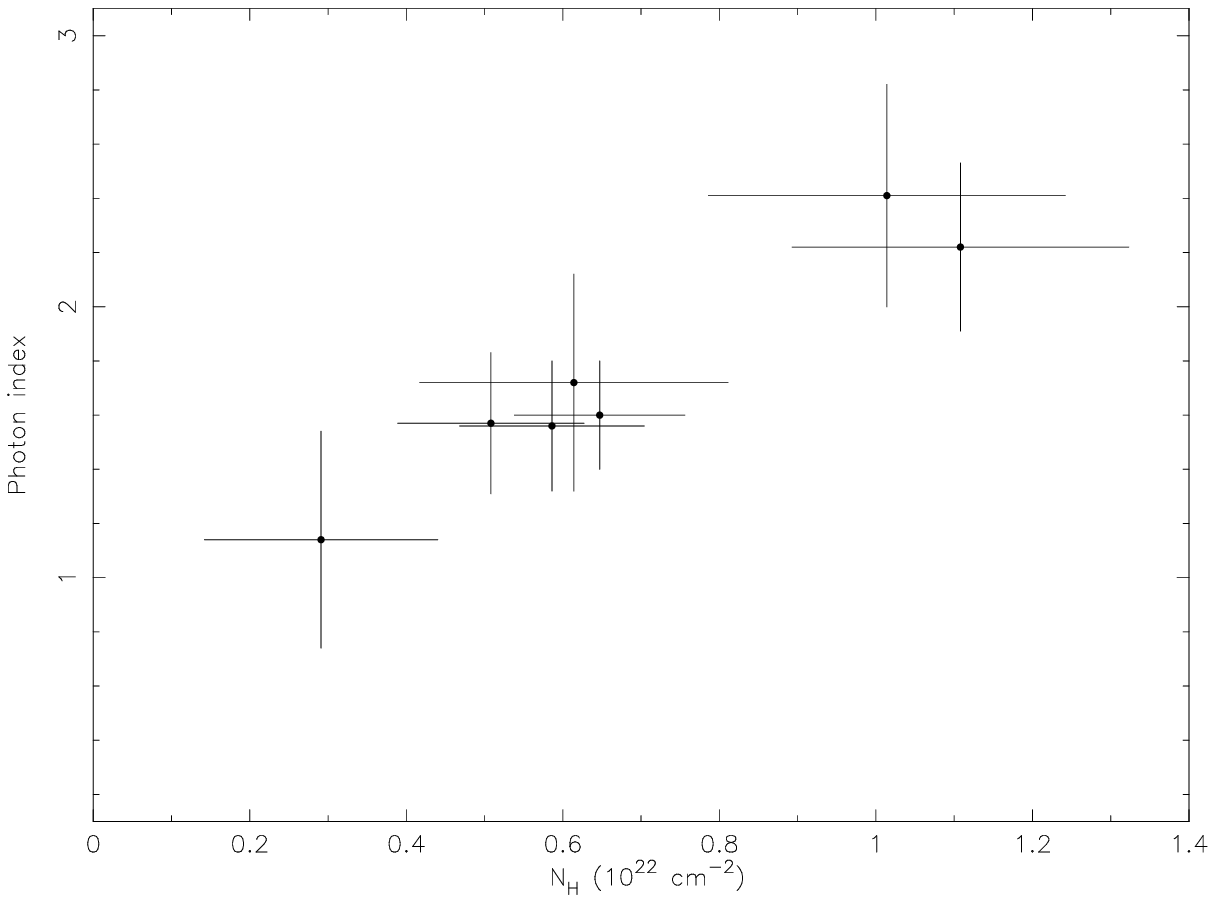,width=3in,height=2.45in}
}
\centerline{
\begin{minipage}[h]{3.5in}
{\small Figure 4: {Plot of power-law photon index ($\alpha$) against absorption
column density ($N_H$). A positive correlation can be seen. \\}}
\end{minipage}
}}

\section{Discussion}

We analyzed the \chandra\ ACIS-S X-ray spectra of four BHXN in
quiescence by fitting the spectra to simple one component models
(power-law, thermal Bremsstrahlung, Raymond-Smith, blackbody)
including interstellar absorption.  While the statistics afforded by
the \chandra\ data surpass that previously available, they are still
inadequate to rule out any of these simple model, except for the
blackbody model in the case of V404 Cyg.  There is some weak
additional evidence against a few other models: For \nmon\ the
Raymond-Smith and blackbody models imply unlikely values of $N_H$
which are lower than the optically-determined values.  The same is
true for \gro\ and \xte\ in the case of the blackbody model.  On the
other hand, the thermal bremsstrahlung model provides a good fit to
the data in all cases; however, the physical interpretation of this
model is unclear (see Christian \& Swank 1997).  The model which does
fit well in all cases and which has a straightforward physical
interpretation is the power-law model with a photon index of $\sim 2$.
This slope is consistent with the spectra expected for an ADAF.

Bildsten \& Rutledge (2000) suggest that much of the X-ray flux
observed from quiescent BHXN may be produced by a rotationally
enhanced stellar corona in the secondary star, as seen in
tidally-locked binaries such as the RS CVn systems.  Lasota (2000) has
criticized this view, suggesting instead that the physically smaller
secondaries of CVs provide a better analog, and that in this case the
expected coronal emission is far below that seen in quiescent BHXN.

The coronal hypothesis of Bildsten \& Rutledge (2000) makes two clear,
testable predictions.  First, that $L_X < 10^{-3} L_{bol}$, and
second, that the spectrum of the quiescent BHXN should be similar to
that of a stellar corona, i.e., well represented by a Raymond-Smith
model with $kT < 1.4$~keV, (Dempsey et al. 1993). The luminosity and
spectral evidence available for five of the six BHXN observed by
\chandra\ rule strongly against these hypotheses, as detailed below.
Note that we do not include in this discussion a seventh BHXN observed
with \chandra\, 4U~1543-47, (see G01) because it contains a fully
radiative secondary (Orosz et al. 1998) and is not expected to possess
an X-ray corona.  Consequently, this system is irrelevant to the
present discussion.

Figure~5, which is adapted from Bildsten \& Rutledge (2000), compares
the quiescent fluxes of BHXN with the predictions of the coronal
model.  The quiescent flux of GRO J0422+32 exceeds the maximum
prediction of the coronal model by a factor of $\sim 60$, and V404~Cyg
exceeds this limit by a factor of $\sim 40$.  \xte\ exceeds the
coronal limit by a factor $\sim 50$ (or $\sim 400$) for the highest
(or lowest) $L_{bol}$ computed in section 4.4.  However, the
luminosity of \xte\ should be treated with caution because a
mini-outburst occurred 120~d after this observation (see Tomsick et
al. 2001).  This situation is very similar to the case of the \asca\
observation of \gro\ made between two outbursts which gave a high
value of the luminosity (see Section 4.3).  Finally, A0620--00 is a
factor of $\sim 5$ above the coronal prediction, which may be a
significant discrepancy since the prediction corresponds to the
maximum likely level of coronal emission.

Turning to the spectral evidence, we find herein that the
X-ray spectra of V404 Cyg, A0620-00, GRO J1655--40 and \xte\ are harder
(equivalently hotter) than typical spectra of stellar coronae.  The
average temperature for these sources as determined from the $N_H$
free (fixed) fits to Raymond-Smith models is 7.4~keV (10.1~keV).  
The average of the 90\% lower limits to the temperatures is
$>4$~keV (or $>5.24$~keV from the $N_H$ fixed fits). 
Coronal sources are often fit by Raymond-Smith models with two
separate temperature components.  The average of the higher of
these temperatures has a value of 1.4~keV (Dempsey et al.  1993).  
Thus, in the four systems for which the data
are of sufficient quality to allow us to measure the X-ray spectrum,
the temperature is $\sim 5$ to 7  times higher than the 
higher temperature typically seen from stellar corona. 

\vbox{
\centerline{
\psfig{figure=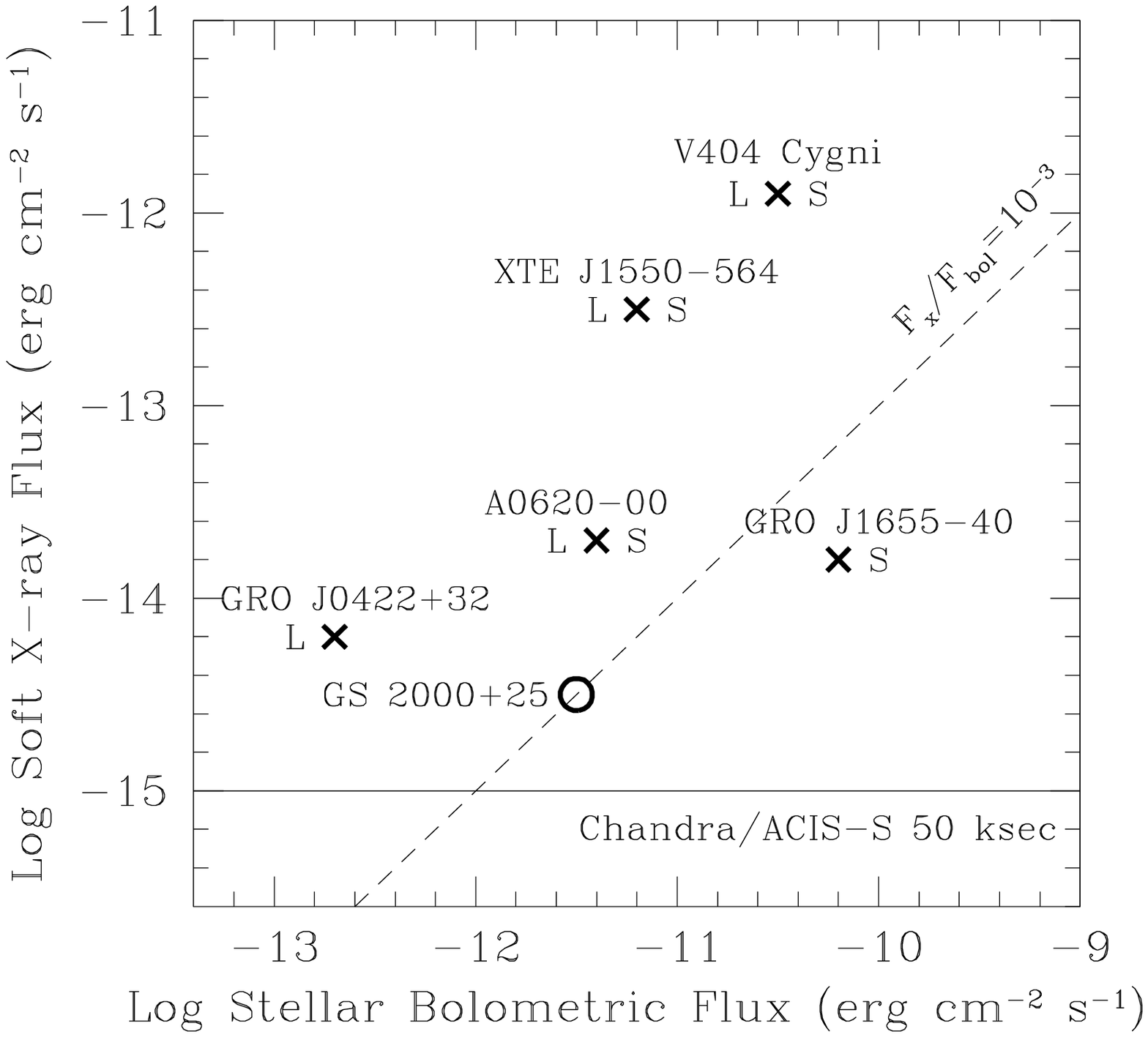,width=3.5in}
}
\centerline{
\begin{minipage}[h]{3.5in}
{\small Figure 5: {Quiescent X-ray and Bolometric Fluxes of BHXN, after Bildsten
\& Rutledge (2000).  X-ray fluxes are as reported herein or from G01,
but in all cases converted to 0.4--2.4~keV emitted fluxes (note that
Figure~2 of Narayan, Garcia and McClintock (2001) plotted a
0.5--10.0~keV flux for V404 Cyg but agrees with this plot in all other
respects).  In cases
where the spectrum cannot be determined, we have assumed $\alpha =
2$. Bolometric fluxes are
from Bildsten \& Rutledge (2000) or as reported herein.  An ``X''
indicates that the X-ray flux is unlikely to be due to a stellar
corona, An L (S) indicates that it is the X-ray luminosity (spectrum)
that argues against this coronal hypothesis.  Based on the data
herein, in only one (GS 2000+25) of the six BHXN studied could the
corona of the secondary produce a signficant part of the detected X-ray 
flux. \\}}
\end{minipage}
}}

Thus the combination of spectral and luminosity information argue
against a coronal source for the quiescent luminosity in 5 out of the
6 cases for which the coronal mechanism is potentially relevant
(i.e. excluding 4U~1543-47).  Only in the case of GS~2000+25, where we
are unable to determine a spectrum due to the very low number of
counts, is it possible that coronal emission from the secondary
dominates the quiescent luminosity.

During strong flares, stellar coronae are occasionally seen at
temperatures higher than the 1.4~keV average value quoted above.  For
example, a ``super-hot giant flare'' from Algol was seen to have a
peak temperature of 12.37~keV (Favata \& Schmitt 1999).  In this
regard, it is important to note that both A0620-00 and \gro\ were
observed with \chandra\ at {\it lower} luminosities than in previous
quiescent observations.  Therefore it is unlikely that these two
systems were in a flaring state during our observations.

Either the secondaries in BHXN have coronae unlike those seen before,
or the source of the quiescent luminosity is not coronal.  This is not
to say that these secondaries do not have X-ray emitting corona, but
merely that the luminosity from such a corona is swamped by the accretion
luminosity even during quiescence.  An obvious point to note is the
following.  Emission from a stellar corona will contribute at some
level to the quiescent X-ray luminosity.  If in a few cases this level
is significant, then the accretion luminosities of the black holes
must be even lower than our estimates and the argument for event
horizons would be further strengthened.

It is worth noting that the five BHXN for which coronal emission is
ruled out cover the full range of orbital period and stellar
bolometric flux.  It therefore seems unlikely that there is some
particular region of parameter space where the coronal model applies.
In comparison, the ADAF model is consistent with all the observations,
covering the full parameter space (Narayan et al. 1996, 1997a, 2001; 
Lasota
2000).

Results of this paper further constrain the required ADAF
model. Quataert \& Narayan (1999) proposed that significant mass can
be lost to an outflow/wind in ADAF models. They also predicted the
spectral shape for ADAF models with and without winds for \cyg. Our
observations indicate that the power-law photon indices of \cyg,
\nmon, \gro\ and \xte\ are consistent with $\alpha\sim2$. Therefore,
models in which Comptonization dominates are favored (Narayan et
al. 1997a); strong-wind models become unlikely unless $\delta$ (the
fraction of the turbulent energy which heats the electrons) is large
enough (Quataert \& Narayan 1999). ADAF models also predict line
emission in X-ray spectra (e.g. Narayan \& Raymond 1999). We set an
upper limit on the equivalent width of any line feature between 6.4--7
keV for \cyg\ and it is much higher than the theoretical prediction
even for model with winds. A larger collecting area instrument like
\xmm\ is needed to study this kind of feature.  Recent {\it RXTE} and
\chandra\ observations of \xte\ also suggest that the ADAF model can
explain the quiescent X-ray emission, although it does not explain all
the behavior observed at other wavelengths (Tomsick et al. 2001).  
The similarity of the
quiescent spectra of \cyg, \xte, \gro, and A0620-00 found herein
suggests that they may all be described by a similar ADAF model.
Detailed broadband spectral modeling of these systems should be
considered in order to further constrain the models.

\vbox{
\centerline{
\psfig{figure=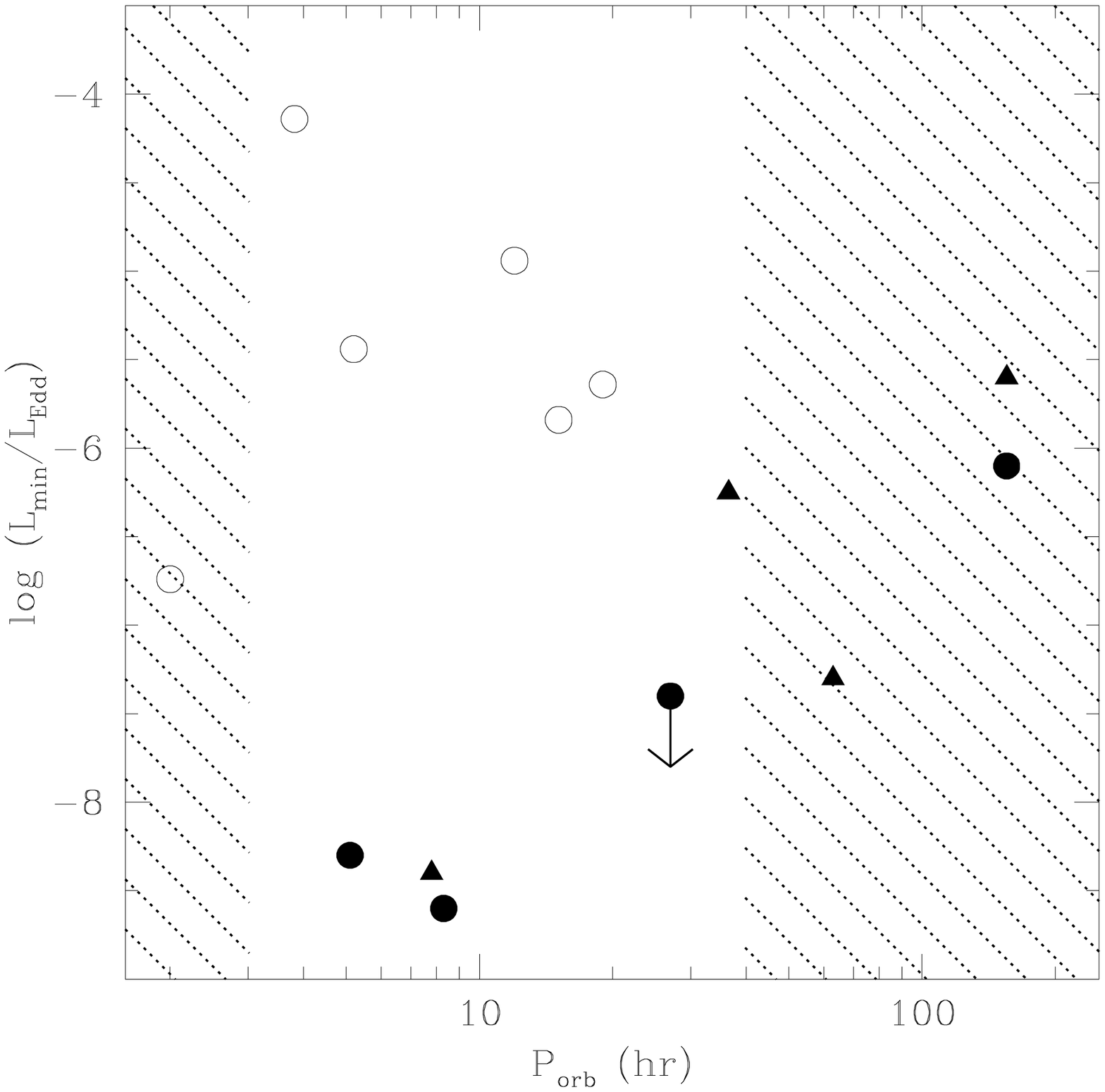,width=3.3in}
}
\centerline{
\begin{minipage}[h]{3.5in}
{\small Figure 6: {Quiescent luminosities of BHXN (filled circles and triangles) 
and NSXN
(open circles) after G01. Data points in triangle are from the results
of this work. Only the lowest quiescent detections (except for \cyg\
which we show both the lowest detection and the luminosity derived
here) or \chandra\ upper limits are shown. The non-hashed areas 
represent
common orbital periods for BH and NSXN. The BHXN shown are, from left
to right: GRO\,J0422+32, \nmon, GS\,2000+25, 4U\,1543--47, \xte, \gro\
and \cyg. \\}}
\end{minipage}
}}

The sources discussed in this paper have a wide range of
luminosities. \cyg\ is the brightest quiescent BHXN in our sample,
with a 0.3--7 keV luminosity of $\sim 5\times 10^{33}$ \lum. In our
\chandra\ observations the source was somewhat more luminous than in
previous quiescent observations in which the luminosity was about
$10^{33}$ \lum\ (see Table 1). Wagner et al. (1994) reported that
\cyg\ exhibited a decrease in intensity by a factor of 10 in $<$ 0.5
day, while our \chandra\ observations showed a factor of 2 variability
in a few ksec. Wagner et al. (1994) also found that there may have
been a factor $\sim$ 2 variability on timescales of $\sim 30$ minutes
in the highest intensity bins for the \rosat\ observations. Thus \cyg\
in quiescence shows variability in X-rays on both short-term (a few
ksec) and long-term (years) timescales.  Significant X-ray variability
in quiescence was also seen in 4U\,1630--47 (Parmar et al. 1997),
\nmon\ (Asai et al. 1998; Menou et al. 1999; also Table 1) and
GX\,339--4 (Kong et al. 2000). \cyg\ and GX\,339-4 are similar in some
respects: for example, their quiescent X-ray luminosities are
comparable (Kong et al. 2000, 2002) and GX\,339--4 has also been
observed to undergo X-ray variability by a factor of 3 during its
quiescent or `off' state (Kong et al. 2002).  Thus, variability in the
quiescent state is common, which suggests that BHXN in quiescence are
not totally turned off. We note that XTE\,J1550--564 also varied in
luminosity by factor of $\sim 2$ during the two \chandra\ observations
in quiescence (Tomsick et al. 2001); only \cyg\ and GX\,339--4 have a
quiescent luminosity higher than XTE\,J1550--564.

In Figure 6, we plot the Eddington-scaled luminosities (based on the
best-fit power-law model) as a function of orbital period $P_{orb}$;
this is an update of the same plot from G01. For the mass of \xte\, we
assumed $M=10.6 M_{\odot}$ (Orosz et al. 2001); the distance to \xte\
is estimated to be 2.5--6.3 kpc
(e.g. S$\acute{a}$nchez-Fern$\acute{a}$ndez et al. 1999; Orosz et
al. 2001) and we have adopted an average distance of 4 kpc.  We note
that for the three long orbital period systems (\cyg: 6.47 d, \gro:
2.6 d and \xte: 1.55 d), the quiescent luminosity is higher than for
the other systems (Figure 6). This implies that the accretion rate in
these systems is higher than for the others according to the ADAF
model (Narayan et al. 1997a; Menou et al. 1999).  It is not clear if
there is a positive correlation between the luminosities and orbital
periods (see Figure 6); a larger sample of long orbital period systems
is required to study this correlation.

In summary, we note that our results confirm the prediction of Lasota
(2000), who previously pointed out that X-ray emission from a
quiescent BHXN is unlikely to come from a stellar coronae; instead he
argues that the emission is due to an ADAF.  Based on this model,
Lasota (2000) predicted fluxes similar to those reported herein.
Moreover, he pointed out that detection of GRO\,J0422+32 by \chandra\
would rule out the coronal model, and such a detection has been made
(G01).  However, our \chandra\ spectra are able to rule out only a few
of the simple, one-component spectral models we fit to the data.  With
its larger collection area, observations with \xmm\ should be able to
do a significantly better job of constraining the source spectra.
Additionally, we note that V404~Cyg is variable on a few ksec
timescale, so simultaneous optical and X-ray observations may shed
substantial light on the quiescent accretion processes in this source.

\begin{acknowledgements}
AKHK was supported by a Croucher Fellowship. JEM was supported in part
by NASA grant GO0-1105A. MRG acknowledges the support of NASA LTSA
Grant NAG5-10889 and NASA Contract NAS8-39073 to the CXC. The HRC GTO
program is supported by NASA Contract NAS-38248.
\end{acknowledgements}


\end{document}